\documentclass[aps,amsmath,amssymb,showpacs,twocolumn,prl]{revtex4-1}
\usepackage{graphicx}
\usepackage{dcolumn}
\usepackage{bm}

\newcommand{\be}{\begin{equation}}
\newcommand{\ee}{\end{equation}}
\newcommand{\Tr}{\mathrm{ Tr }}

\newcommand{\normfig}{0.90\linewidth}

\newcommand{\ket}[1] {\vert{#1}\rangle}

\begin{document}

\title{Renormalization group contraction of tensor networks in three dimensions}
\author{Artur Garc\'ia-S\'aez and Jos\'e I. Latorre}
\affiliation{Departament d'Estructura i Constituents de la Mat\`{e}ria, Universitat de Barcelona, 08028 Barcelona, Spain}

\begin{abstract}

We present a new strategy for contracting tensor networks in arbitrary geometries. 
This method is designed to follow as strictly as possible the renormalization group philosophy, by first 
contracting tensors in an exact way and, then, performing a controlled truncation of the resulting tensor. 
We benchmark this approximation procedure in two dimensions against an exact contraction. 
We then apply the same idea to a three dimensional system. 
The underlying rational for emphasizing the exact coarse graining renormalization group step prior to truncation is related to monogamy of entanglement.

\end{abstract}

\maketitle

Computing analytically the properties of a quantum model is, in general, not possible. 
It is then necessary to resort to classical simulations that rely on approximation techniques. 
From a quantum information perspective, the 
presence of a small amount of entanglement in the system has been identified
as the key ingredient that allows for efficient classical simulations.
This has been exploited in one dimension ($1$D) in a series of algorithms based on a description
of the system as a tensor network. Following this approach, condensed matter
systems can be simulated using Matrix Product States (MPS, \cite{Werner,Ostlund}) 
as the testbed for the variational procedure of the density-matrix 
renormalization group (DMRG, \cite{dmrg}) to study ground states of local $1$D Hamiltonians. 
Beyond the $1$D setting, the computation of physical magnitudes 
using tensor networks is limited by the numerical effort necessary 
to perform the contraction of the tensor network, {\emph i.e.} to sum all its
indices. The efficiency for performing this task is limited  
by the area law scaling of the entanglement entropy in the system \cite{vdrk,areaR,riera,cardy}.
To overcome this problem, several strategies
aim at finding the best possible approximation to the contraction of tensor
networks after identifying the relevant degrees of freedom of the system \cite{vidal,faithf,luca}.

Let us briefly recall the key elements of the tensor network representation. 
Given a quantum state of $N$ particles $\ket{\psi} = \sum c^{i_1,\ldots,i_N} \ket{i_1\ldots i_N}$ 
its coefficients can be represented as a contraction of local tensors
 $c^{i_1,\ldots,i_N}={\rm tr} (A^{1,i_1}\ldots A^{N,i_N})$,  where each local tensor $A^j$ carries
a physical index $i_j$ and ancillary indices (which are not written) that get contracted according
to a prescribed geometry. The rank of these ancillary indices, that we shall call $\chi$,
controls the amount of entanglement which is captured by the tensor representation.
If the tensors are simple matrices on a line, the tensor
network is called MPS \cite{Werner,Ostlund}. 
Other possible geometries are regular squared grids in any dimensions
that correspond to PEPS \cite{VC04}, and tree-like structures that go under the name of TTN \cite{ttn} and MERA
\cite{VidMERA}.

In this letter we propose a new strategy to
contract tensor networks in general geometries, that we shall illustrate in detail
for PEPS in 2D and 3D.
The method is based on following as strictly as possible the renormalization group (RG)
philosophy. First, an exact contraction of a set of local tensors is performed that 
produces a coarse grained tensor of larger rank. Subsequent contractions would made the rank 
of the effective tensors to scale 
following a law dictated by the geometry of the tensor network. For instance, in the
case of PEPS the rank of the coarse grained tensors would grow following an area law.
It is then necessary to perform a truncation that faithfully retains only relevant
degrees of freedom. To achieve this truncation, our method makes a series of 
Schmidt decompositions that cast the relevant information onto a
renormalized tensor. 

The strategy we present here differs from a previous 
renormalization group inspired proposal \cite{levin,wen,xiangPRB}. There, 
the original tensor is first truncated and then contracted efficiently.
Instead, we first contract exactly tensors at a larger numerical cost, and then truncate.
This procedure can be made exact in $1$D \cite{vclrw}. In more dimensions,
this idea entails a trade off between numerical computation speed and precision.
Our proposal relies on the idea that in higher dimensions, {\sl e.g.} $3$D, 
monogamy of entanglement makes every degree of freedom to have a reduced amount
of entanglement with each neighbor. Long distance correlations emerge from
the multiplicity of possible paths connecting local degrees of freedom. 
Therefore, a tensor network
with small rank  $\chi$ is already a good approximation to a $3$D system. 
The fact that good tensor network representations only need
a small $\chi$ makes viable our
proposal for exact contractions followed by truncation.

Let us recall that, in 2D settings, some variants of contraction schemes for PEPS have been analyzed
\cite{VC04,levin,wen,xiangPRB}. To our knowledge, PEPS renormalization methods have not been applied to 3D systems, 
which nevertheless have been studied using cluster states \cite{anders} or string bond states \cite{string3D}.
For $3$D classical systems other renormalization algorithm have been proposed \cite{nish1,nish2}.


\emph{RG contraction in $2$D.---} 
Let us illustrate our method for contraction of tensor networks with the 
computation of the norm of a state in $2$D. 
We start by considering the norm of the original state $|\psi\rangle$,
that is the
folded lattice made of tensors of the form
 $E^{(m,n)} = \sum_i A^{*(m,n)i}\otimes A^{(m,n)i} $, where 
$(m,n)$ labels a position in the $2$D lattice and physical indices $i$ at this point 
have been summed over. 
Each tensor $E^{(m,n)}$ has four further  ancillary indices 
of  rank $\chi^2$ (not written explicitly) pointing to its $2$D neighbors.
The global contraction of the $2$D tensor network corresponds
to the computation of the norm
$\langle \psi|\psi \rangle = \Tr\left( E^{(1,1)} \ldots E^{(N,N)}\right)$,
where $\Tr$ is a generalized trace that contracts all ancillary indices.

The stages in our contraction strategy go schematically as follows.
We start with the tensor $E^{(m,n)}$ which carries 4 indices of rank $\chi^2$.
We first contract four adjacent tensors to build a plaquette
$P^{(m,n)}=E^{(m,n)} E^{(m,n+1)} E^{(m+1,n)} E^{(m+1,n+1)}$.
Now $P^{(m,n)}$ carries a total of 8 open indices of rank $\chi^2$,
showing the area law increase of indices naturally associated to the contraction
of tensors in a 2D square lattice. 
This plaquette tensor  will be approximated by a renormalized tensor 
on a coarse grained lattice
$\tilde{E}^{(m,n)}\approx P^{(m,n)}$,
where we shall truncate to a tensor of 4 indices of rank $\chi^2$.
The global contraction symbolically reads
\be \label{eq:Z}
\langle\psi|\psi\rangle=
\Tr\left(\{ E \} \right)=\Tr \left( \{ P \}\right) \approx \Tr\left( \{\tilde{E} \}\right),
\ee
which can be graphically represented as
\begin{align}\nonumber
\includegraphics[width=0.4\textwidth]{dg1.eps} 
\end{align}
where tensors $E$ are represented by nodes connected to its neighboring tensors 
with lines that represent ancillary indices. 
After a renormalization step, we recover the same lattice structure with 
renormalized tensors.

Let us now discuss in more detail each step in the renormalization
group contraction we have just sketched.
Though computationally expensive, the initial exact 
contraction $\{ E \} \to \{ P \}$
has two main advantages. First, an exact contraction of
the tensors in a plaquette retains 
all degrees of freedom, making
this method specially suitable for systems with frustration at the plaquette scale.
Furthermore, by an appropriate renormalization of the virtual bonds of the $P$ tensors 
we recover the original lattice structure, modulo an increase in the rank
of ancillary indices. When completing the strategy with a truncation, it will then
be possible to perform
a systematic iteration of the same procedure, resulting in the global
contraction of the tensor network. The method is thus a way to scape
from the area law restriction at the cost of a truncation of the renormalized
tensors.

Once we have constructed the renormalized tensor $P^{(m,n)}$, we
need to perform a controlled truncation. Let us 
modify the notation to make this step clearer, by dropping the site label and introducing
explicitly the ancillary indices $P^{(m,n)}= P_{\alpha_1 \alpha_2 \alpha_3\alpha_4}$.
Each ancillary index $\alpha$ is the result of combining the two ancillary indices
that were attached to tensors $E$ pointing towards an adjacent plaquette.
To be concrete, we call $\alpha_1$ the index connecting the plaquette to its upper neighbor.
This index is the combination of two indices from the original tensors $\alpha_1=\beta_1 \otimes \beta'_1$.
Thus, each $\alpha$ is a combined index of rank $\chi^4$.
We then select the index $\alpha_1$
and separate it from the rest of indices of the plaquette by means of a 
singular value decomposition
\be
\label{plaquettesvd}
P_{\alpha_1 \alpha_2 \alpha_3 \alpha_4} = U_{\alpha_1\mu}
\lambda^\mu V_{\mu\alpha_2\alpha_3\alpha_4}
\ee
where $U$ and $V$ are unitary matrices and $\lambda^\mu$ are the eigenvalues
in the decomposition. It is convenient to include the 
squared root of the eigenvalues into the unitary matrix $U_{\alpha_1\mu}$ as follows
\be
W_{\alpha\mu}=U_{\alpha\mu}\sqrt{\lambda^\mu}.
\ee
We repeat this process for each index in order to obtain the four matrices $W_1,W_2,W_3$ and $W_4$
as follows
\begin{align}\label{decompU}\nonumber
\includegraphics[width=0.48\textwidth]{dg2.eps} 
\end{align}
After these four successive decompositions, we can construct a new tensor
 $\tilde{P}=W^{-1}_1 W^{-1}_2 W^{-1}_3 W^{-1}_4 P$, where we have used
 pseudo inverse matrices.


We proceed now to perform a renormalization of the plaquette $P$ using the set of tensors $W$s. 
The key observation here is that, along each direction, the plaquette is surrounded by other plaquettes decomposed in a similar way. 
We assume that a plaquette $P = W_1W_2W_3W_4 \tilde{P}$ has neighboring
unitaries $W'_1,W'_2,W'_3$ and $W'_4$. The renormalization consists on the truncation of the degrees of freedom 
between plaquettes through the unitaries $W$ and $W'$.
We first take matrices $W_1$ and $W'_3$ to form tensor $R_{3,1} = W'_3 W_1$ which can be decomposed using a SVD into
$R_{3,1} = \tilde{U}\Sigma \tilde{V}$. The truncation corresponds to only keeping 
the largest $\chi^2$ eigenvalues from matrix $\Sigma$ to generate the new decomposition
   $R_{3,1} \approx I_3 I_1$, where each matrix $I$ is the projection to $W$ onto the
relevant eigenvalue subspace.
We repeat this procedure for each neighboring pair  $W$ and $W'$. We then apply the four matrices $I_k$ to $\tilde{P}$ to obtain our final truncation
$ \tilde{E}=I_1 I_2 I_3 I_4 \tilde{P}$
\begin{align}\nonumber
\includegraphics[width=0.48\textwidth]{dg3.eps} 
\end{align}
This truncation strategy that delivers the renormalized tensor $\tilde{E}$
can be represented as
\be\nonumber
P=\Tr(EEEE)=W_1W_2W_3W_4\tilde{P} \approx I_1I_2I_3I_4 \tilde{P}=\tilde{E}
\ee
and reduces the size of the plaquette tensor to  $\chi^2$ for each bond.

The procedure described above is computationally demanding compared to variational methods
or other renormalization techniques. In order to make the singular value decomposition in
Eq. \ref{plaquettesvd} it is necessary to make the contraction of
a plaquette with its own adjoint
\begin{align} \label{TT}
\includegraphics[width=0.48\textwidth]{dg4.eps} 
\end{align}
to obtain each of the $W_k$. This strategy outperforms
the naive use of singular value decomposition algorithms.

As mentioned above, previous proposals for the renormalization of tensors networks 
rely on a decomposition of the local tensors before
producing a truncation \cite{levin,wen,xiangPRB}. Such a decomposition allows to achieve a higher 
bond dimension after the original truncation. Other approaches work with square plaquettes that include the physical indices \cite{sandvik}. 
It is also important to notice that the method proposed here is not a variational procedure,
though it remains numerically stable.

\emph{Validation in $2$D.--} We validate our approach showing results for the renormalization of a 
tensor network representing  the ground state of the Ising Hamiltonian with a transverse magnetic field
in a square lattice
\be \label{ham}
H = -\sum_{ \{i,j\} } \ \sigma_i^z \sigma_j^z + h \sum_i \sigma_i^x .
\ee 
where $\{i,j\}$ are neighboring sites on finite or infinite lattices. 
This model displays a phase transition at a critical field $h_c\approx 3.04$ 
for the infinite square lattice.

The preparation of the tensor network can 
be performed using an imaginary time evolution \cite{MPDO,VC04}, 
$|\psi_f\rangle = e^{-Ht}|\psi_0\rangle $.  We first proceed
to use a Trotter approximation to this Euclidean evolution.
At each step we have to apply the evolution operator $T$ to a pair
of neighboring tensors $\Theta_{\alpha,\beta}^{i',j'}= T_{i,j}^{i',j'} [A^i_{\alpha,\mu}]_k [A^j_{\mu,\beta}]_{k+1}$
and truncate the new tensors $[\tilde{A}^i_{\alpha,\mu}]_k$ and $[\tilde{A}^j_{\mu,\beta}]_{k+1}$
to the lattice bond dimension. 

\begin{figure}
\includegraphics[width=\normfig]{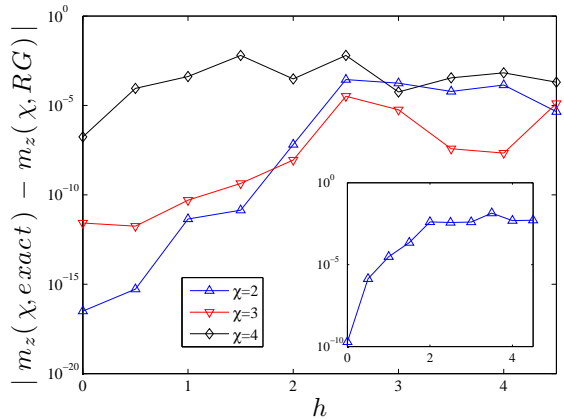}
\caption{For an Ising model with transverse field in a finite 2D square lattice of size $6\times 6$, 
we plot the error of the magnetization  between the exact contraction and our renormalization group
contraction of the same PEPS, that is $m_z(\chi,exact)-m_z(\chi,RG)$, for different values of $\chi$. 
Inset: Same for a $12\times 12$ lattice  with $\chi=2$.}\label{fig1}
\end{figure}

In order to validate our strategy,
we show in Fig.\ref{fig1} the error of the magnetization $m_z$ in a finite square lattice 
of size $6\times 6$ and $12\times 12$, with our RG contraction 
against the exact contraction of the very same tensors, that is
$|m_z(\chi,exact)-m_z(\chi,RG)|$. The error introduced by the RG contraction is of the order
of $10^{-3}$ at the critical point. Let us note that this error is smaller for $\chi=2$ since 
the RG casts $\chi^{16}$ numbers to $\chi^8$. It is thus expected that the 
faithfulness of RG is better for small $\chi$. 
Obviously, the $\chi=4$ tensor is in itself a better representation of the state.

\emph{$3$D quantum systems.--}
The contraction strategy for tensor networks we have presented in $2$D can
be extended for a $3$D square lattice.
The renormalization is performed over plaquettes of eight tensors forming a cube.
We use a singular value decomposition to decouple at each step four ancillary indices, and obtain the unitaries
corresponding to each orthogonal direction. This set of six matrices $W_k$ is combined with unitaries $W'_k$
associated to neighboring plaquettes, to produce six  truncated matrices $I_k$.
We perform a similar operation along each direction to recover the renormalized tensors with bond dimension $\chi^2$.
To overcome limitations in computational
resources, we use an operation similar to the trick in \ref{TT} and
obtain each of the unitaries $W_i$. 
For a single direction, this renormalization is represented as
\begin{align}
\includegraphics[width=0.4\textwidth]{dg5.eps} 
\end{align}

It is possible to prepare the original state with a tensor network using
some specific $\chi$. Yet, after a first contraction is made, we can retain
renormalized tensors with a larger effective $\chi_{eff}$. We have
checked that for $\chi=2$, stable results are obtained for $\chi_{eff}\sim 5$.

\begin{figure}
\includegraphics[width=\normfig]{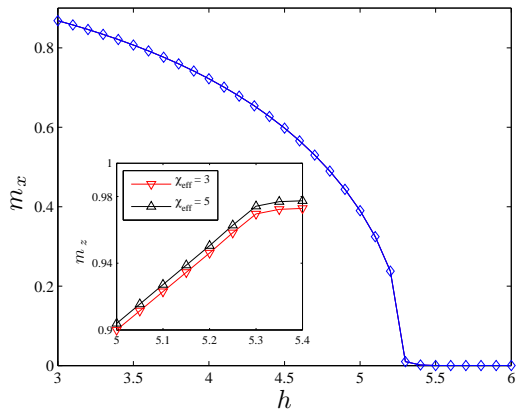}
\caption{Magnetization $m_z$ vs. local field $h$ for the Ising model with transverse field in 
an infinite 3D square lattice. The critical point is found at
$h_c\approx 5.29$. The rank of the tensor network is $\chi=2$.
Inset: Magnetization $m_x$ vs. $h$ for different values of $\chi_{eff}$.}\label{fig2}
\end{figure}

We can now apply our approach to 3D systems to compute the magnetization of Ising model
Eq.\ref{ham} for an infinite 3D square lattice, by iterating the above procedure.
The state is prepared using a minimal environment to stabilize the Trotter evolution.
The results for the magnetizations 
$m_z$ and $m_x$ for $\chi=2$ are presented in Fig.\ref{fig2}. A quantum phase
transition is detected at a critical point located around $h_c\simeq 5.29$ (series expansions
detect a critical point at $h_c = 5.14$ \cite{3Dser}).
We have also computed the equivalent phase transition in 2D, using again
only $\chi=2$ tensors. In the 2D case, our RG contraction finds a transition at
$h_c\simeq3.25$ to be compared with the result $h_c=3.04$ coming
from other more precise methods. This hints at the fact that RG contraction is a better
approximation in 3D than in 2D.

\begin{figure}
\includegraphics[width=\normfig]{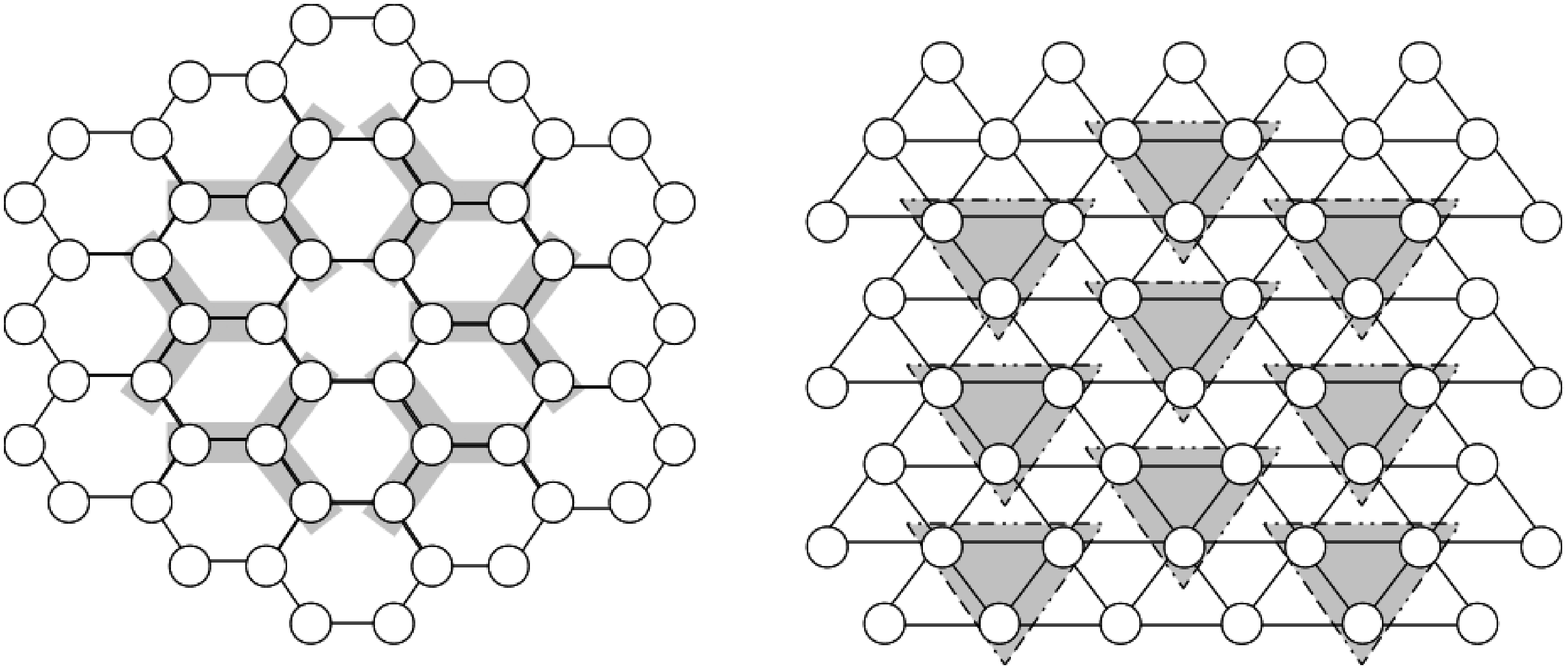}
\caption{Renormalization of hexagonal and triangular lattices,
forming plaquettes with the tensors inside the shaded areas.}\label{fig3}
\end{figure}

\emph{Conclusions.---} 
We have presented a novel, RG inspired strategy to contract
tensor networks that delivers good results in a 3D simulation
of the phase transition in the quantum Ising model. 
Our scheme can be extended to 
hexagonal and triangular lattices, where a renormalization step 
can be performed choosing the plaquettes to be contracted as in Fig.\ref{fig3}
so as to  recover a rescaled version of the original lattice.
This opens the possibility of studying frustrated models in 3D 
using PEPS technology.

We acknowledge financial support from MICINN (Spain), Grup de Recerca Consolidat (Generalitat de Catalunya), 
QOIT Consolider-Ingenio 2010 and ICREA-ACAD\`EMIA.


\begin{thebibliography}{10}

\bibitem{Werner} M. Fannes, B. Nachtergaele and R. F. Werner, Commun. Math. Phys. {\bf 144}, 143 (1992).
\bibitem{Ostlund} S. \"Ostlund and S. Rommer, Phys. Rev. Lett. {\bf 75}, 3537 (1995).

\bibitem{dmrg} S. White, Phys. Rev. Lett. {\bf 69}, 2863 (1992).

\bibitem{vdrk} G. Vidal, J. I. Latorre, E. Rico and A. Kitaev, Phys. Rev. Lett. {\bf 90}, 227902 (2003).
\bibitem{cardy} P. Calabrese and J. Cardy, J. Stat.Mech. P06002 (2004).
\bibitem{riera} A. Riera and J. I. Latorre, Phys. Rev. A {\bf74}, 052326 (2006).
\bibitem{areaR} J. Eisert, M. Cramer and M. B. Plenio, Rev. Mod. Phys. {\bf82}, 277 (2010).

\bibitem{vidal} G. Vidal, Phys. Rev. Lett. {\bf91}, 147902 (2003).
\bibitem{faithf} F. Verstraete and J.I. Cirac, Phys. Rev. B {\bf73}, 094423 (2006).
\bibitem{luca} L. Tagliacozzo, T. R. de Oliveira, S. Iblisdir and J. I. Latorre, Phys. Rev. B {\bf 78}, 024410 (2008).

\bibitem{VC04} F. Verstraete and J. I. Cirac, arXiv:cond-mat/040766 (2004).

\bibitem{ttn} L. Tagliacozzo, G. Evenbly and G. Vidal, Phys. Rev. B {\bf80}, 235127 (2009).
\bibitem{VidMERA} G. Vidal, Phys. Rev. Lett. {\bf 101}, 110501 (2008).

\bibitem{levin} M. Levin and C. P. Nave, Phys. Rev. Lett. {\bf 99}, 120601 (2007).
\bibitem{wen} Z. C. Gu, M. Levin and X. G. Wen, Phys. Rev. B {\bf 78}, 205116 (2008).
\bibitem{xiangPRB} H. H. Zao, Z. Y. Xie, Q. N. Chen, Z. C. Wei, J. W. Cai and T. Xiang, Phys. Rev. B {\bf 81}, 174411 (2010). 

\bibitem{vclrw} F. Verstraete, J. I. Cirac, J. I. Latorre, E. Rico and M. M. Wolf,  Phys. Rev. Lett. {\bf 94}, 140601 (2005).

\bibitem{anders} S. Anders, M. B. Plenio, W. D\"ur, F. Verstraete and H. J. Briegel, Phys. Rev. Lett. {\bf 97}, 107206 (2006).

\bibitem{string3D} A. Sfondrini, J. Cerrillo, N. Schuch and J. I. Cirac, Phys. Rev. B {\bf 81}, 214426 (2010).

\bibitem{nish1} N. Maeshima, Y. Hieida, Y. Akutsu, T. Nishino and K. Okunishi, Phys. Rev. E {\bf 64}, 01670 (2001).
\bibitem{nish2} A. Gendiar and T. Nishino, Phys. Rev. B {\bf 71}, 024404 (2005).

\bibitem{sandvik} L. Wang, Y. Kao and A. W. Sandvik, Phys. Rev. E {\bf83}, 056703 (2011).

\bibitem{MPDO} F. Verstraete, J. J. Garc\'ia-Ripoll and J. I. Cirac, Phys. Rev. Lett. {\bf 93}, 207204 (2004).

\bibitem{3Dser} Z. Weihong, J. Oitmaa and C. J. Hamer, J. Phys. A: Math. Gen. {\bf27}, 5425 (1994).




\end{thebibliography}
\end{document}